\documentclass[aps,prd,twocolumn,showpacs]{revtex4}
\usepackage{graphicx}
\usepackage{latexsym}
\usepackage{amsmath}
\usepackage{amsfonts}
\usepackage{amssymb}

\newcommand{\be}{\begin{equation}}
\newcommand{\ee}{\end{equation}}
\newcommand{\ben}{\begin{eqnarray}}
\newcommand{\een}{\end{eqnarray}}

\begin{document}
\title{Relaxing to a three dimensional brane junction}
\author{D. Bazeia,$^{a}$ F.A. Brito,$^{b}$ and L. Losano$^{a}$}
\affiliation{
\small{$^a${Departamento de  F\'\i sica, Universidade Federal da Para\'\i ba,}
{58051-970 Jo\~ao Pessoa, Para\'\i ba, Brazil}\\
$^b${Departamento de F\'\i sica, Universidade Federal de Campina Grande,}
{58109-970 Campina Grande, Para\'\i ba, Brazil}}}
\begin{abstract}
We suggest a mechanism which leads to 3+1 space-time dimensions. The Universe assumed to have nine spatial dimensions is regarded as a
special nonlinear oscillatory system --- a kind of Einstein solid. There are $p$-brane solutions which manifest as {\it phase} oscillations separating different phase states. The presence of interactions allows for bifurcations of higher dimensional spaces to lower dimensional ones
in the form of brane junctions. We argue this is a natural way to select lower dimensions.
\end{abstract}
\maketitle

{\it I. Introduction.} ---There are recent increasing interest in
high energy physics in understanding why our Universe presents only
three spatial dimensions. Specially in Refs.~\cite{1,2,kr2005} were shown that diluting modes in string gas
\cite{1} and D-brane gas \cite{2,kr2005} cosmology favor
such Universe. In this Letter we report some new ideas on this
question. Following recent idea of space-time being filled with a D-brane gas \cite{2,kr2005}, we suppose
the brane gas dilutes into a nine dimensional system, which we regard as an oscillatory system
with nine non-compact dimensions. The branes manifest as {\it phase
fronts} separating different phase states of oscillations
\cite{front,os}. The system allows for bifurcations to form brane
junctions that we argue to be a natural way to select lower
dimensions. Here, there is the possibility of the ten dimensional
space-time to be filled with a network of junctions of higher
dimensional codimension one branes intersecting orthogonally. In our
set up, junctions of several spatial dimensions $p$ may occur, and
we refer to them as $p$-brane junctions. We argue that gravity,
matter and gauge fields tend to live at $p$-brane junctions with
smaller dimensions. In this sense,
although superstrings predict the Universe should have $9+1$
dimensions, such space-time can be filled with $p$-brane junctions
allowing for the existence of physics at some preferable space-time
with lower dimension. As we discuss below, such space-time can be the
$3+1$ dimensional, as world-volume of 3-brane junctions that fills the ten dimensional space-time,
in accord with Refs.~{\cite{2,kr2005}}. The space selection via brane junctions is
endowed with the phenomenon of gravity, matter and gauge fields
localization on branes \cite{add,rs2} and brane junctions
\cite{nima2000,Csaki99}.

We regard our system as a kind of Einstein solid made out of
$9n$ independent oscillators. The total energy associated with these
vibrational degrees of freedom is simply given in terms of $9n$
decoupled oscillators as $E=1/2\sum\,[{\dot{X}_I}^2+(X_{I})^2$],
$I=1,...,9n$. Following the usual procedure, we consider these
oscillators as normal modes of vibration in a system with nine
spatial dimensions, with the number of oscillators $n\!\to\!\infty$.
The total energy can be written in terms of the energy of a field
theory with $9n$ scalar fields in nine spatial dimensions
$E=1/2\int{{d\,}^9x\sum\,[{\dot{\Phi}_I}^2+(\nabla \Phi_I)^2]}$,
$I=1,...,9n$. This is a conserved energy associated with a field
theory of $9n$ real decoupled scalar fields --- a similar set up
with many decoupled scalar fields describing a large $n$ landscape
of vacua has been consider in \cite{ADK}. As presented above,
the system develops just travelling waves and standing waves
described by the scalar fields $\Phi_I$.


However, our oscillatory system may experience nonlinear phenomena
and exhibit pattern formation appearing as a $p$-brane network. As
it is known, the dynamics of wave amplitudes of a weakly nonlinear
non-relativistic system may be governed by nonlinear Ginzburg-Landau
equations. Specially in periodically forced oscillatory systems,
patterns may be induced by bifurcation, a phenomenon which has been
reported experimentally in \cite{os}. Interestingly, bifurcation can
be controlled by some forcing parameter, which drives stability of
$phase$ oscillations with different phases \cite{front}. Namely, a
``$\pi$ front'' is a kink connecting states with phases of
oscillations that differ by $\pi$. As the forcing strength decreases
the $\pi$ front becomes unstable and decomposes into interacting
$\pi/2$ fronts. In a two dimensional periodically forced oscillatory
system, the amplitude of oscillations satisfies the non-linear
Ginzburg-Landau equations \cite{front}:
$\partial_t\Phi_I=\Phi_I+\frac{1}{2}\nabla^2\Phi_I-\frac{2}{3}\Phi_I^3-\frac{\varepsilon}{2}(\Phi_I^2-3\Phi_J^2)\Phi_I$,
where $I,J=1,2\,(I\neq J)$ and $|\varepsilon|\!<\!1$. Within the context of brane junctions, a $\pi/2$ front configuration in a two
dimensional system is nothing but a $0$-brane junction made out of two orthogonal $1$-branes. Thus in the nine dimensional ``forced''
oscillatory system,  $\pi$ fronts, i.e., $8$-branes are firstly formed. As the forcing strength starts diminishing a sequence of
bifurcations leading $\pi$ fronts to $\pi/2$ fronts takes place in the whole space. In the first bifurcation one has $7$-brane
junctions, in the second one has $7$-brane junctions and $6$-brane junctions, and so on. For a number of scalar fields larger than the
number of spatial dimensions, as a landscape of vacua in string
theory, the bifurcation process leads to 0-brane junctions. Thus if
all $9n$ degrees of freedom  are active, then the lowest dimensional
junctions are $0$-brane junctions. At final stages we may find a ten
dimensional space-time with a network of intersecting $8$-branes
forming $p$-brane junctions of dimensions $p=0,1,2,...,7.$ However,
as we will show below, only six degrees of freedom are activated by
thermal effects, allowing for the least dimensional 3-brane
junctions. This is because the total energy for identical 8-brane
configurations written as
$E_n=1/2\int{{d\,}^9x\sum_{I=1}^{9n}\,[{\dot{\Phi}_I}^2+(\nabla
\Phi_I)^2]}=9n\,T_8\,$Vol$_8$ defines a statistical system with mean
energy $U\!=\!\bar{E}$ that depends on the temperature --- here
$T_8$ and Vol$_8$ are the tension and the volume of the brane.

In this way, the next concern relies on the phenomenon of particle
confinement to such brane junctions. As we mentioned above we argue
that the localization of gravity, gauge and matter fields on
junctions can play the role of relaxing higher spatial dimensions to
just three spatial dimensions. We shall look for phase fronts as
static soliton solutions in the nine dimensional version of the
Ginzburg-Landau equations above. In high energy physics, this is
equivalent to Klein-Gordon equations for soliton solutions in a
field theory with $9n$ real scalar fields in ten dimensional
space-time \ben\label{action}
S=\frac{1}{2}\int{{d}^{10}x\sum_{I=1}^{9n}\left\{\partial_M
\Phi_{I}\partial^M \Phi_{I}-\left[\frac{\partial
W(\Phi)}{\partial\Phi_I}\right]^2\right\}}, \een where we assume the
superpotential is of the form \ben\label{superpot}
W(\Phi)=\sum_{I=1}^{9n} W(\Phi_I); \qquad
W(\Phi_I)=r\,\Phi_I-\frac{1}{r}\,\frac{\Phi_I^3}{3}, \een with
$r=\sqrt{3/2}$. Note we also have left the $\varepsilon$-term that
couples the system to be implemented later
--- this term will be related to stability of junctions.
As has been shown in \cite{ADK} such a coupling is very small for
$n$ very large. The cubic superpotential $W(\Phi)$ will give us a
field theory with $512^n=2^{9n}$ vacua solutions and $8$-brane
solutions that can be joined together to make a $0$-brane junction.
As we have mentioned earlier, there are other junctions as ``edges'', as for example, the $7$-brane junctions which are made
out of intersections of two orthogonal $8$-branes. Our Universe appears to be a $3$-brane junction in this array of different
$p$-brane junctions. Bellow we explore explicitly the brane solutions with few fields involved in the calculations. In a
two-field model, there are only $4=2^2$ vacua at the corners of the plane in the space of fields, and in a three-field model, there
exist $8=2^3$ vacua at the vertices of a cube in the space of fields, and so on. To emphasize the discussions above, we take these
models as concrete examples of junctions \cite{cht} that can occur in different dimensions. In three spatial dimensions, the two-field
model produces only $1$-brane junctions, while in the three-field model, there are $1$-brane junctions and $0$-brane junctions made
out of intersecting $2$-branes.

{\it II. Two-field model}. ---Let us now study in detail the theory (\ref{action}) discussed above in a sector with two real scalar
fields $\Phi_1\equiv u$ and $\Phi_2\equiv v$ given by
\ben\label{lagr} {\cal L}=\frac{1}{2}\partial_\alpha
u\partial^\alpha u+\frac{1}{2}\partial_\alpha v\partial^\alpha v-
\frac{1}{2}\left({\partial_u W}\right)^2-\frac{1}{2}
\left({\partial_v W}\right)^2. \een The superpotential
({\ref{superpot}) now reads \ben
\label{W}W=r\,(u+v)-\frac{1}{r}\left(\frac{u^3}{3}
+\frac{v^3}{3}\right), \een and one-dimensional static soliton
solutions can be found from the first-order equations
\ben\label{BPS} \frac{du}{dx}={r}-\frac{u^2}{r},\qquad
\frac{dv}{dx}={r}-\frac{v^2}{r}. \een These nonlinear equations
which can be easily integrated give us vacuum and soliton solutions. The vacuum solutions are $\bar{u}=\bar{v}=\pm\,r$, and involve a
$\mathbb{Z}_4$ symmetry --- recall $r=\sqrt{3/2}$. These vacua are vertices of a square in the plane $(u,v)$. They form a set of $6$
($4!/2!\,2!$) independent BPS sectors, each one connecting a pair of vertices. Four sectors are at the $edges$ of the square, while the
other two sectors are at the $diagonals,$ which we name edge and 1-diag states. These sectors give us the following soliton (kink)
solutions
\ben\label{edge2}
&&u_a^2=r^2;\:\: v_a^2=r^2\,\tanh^2{(x)},\qquad {\rm edge}\\
\label{diag2} &&u_b^2=v_b^2;\:\: v_b^2=r^2\,\tanh^2{(x)},\qquad {\rm
1-diag}
\een
where $a=1,...,4$ and $b=5,6$. We should recall that we easily found such solutions, because we made the system decoupled by
early setting $\varepsilon\to0$. Indeed, these solutions are independent of each other and we can properly embed them into two
spatial dimensions as domain walls, or 1-branes, separating four vacua. Our point is to show that in the case with ($\varepsilon\neq0$), the
system may {\it favor} the configuration of {\it four edges} solutions (\ref{edge2}) instead of diagonal solutions (\ref{diag2}).
This is the bifurcation phenomenon we have stressed above. In fact,
as we show below, there exist values of $\varepsilon$ which favor
such bifurcation, with the four edges configuration stable.
Such configuration is a stable 0-brane junction. In the {\it thin-wall limit}
\cite{cht}, the junction can be represented by four half-walls
rotated of $90^{\rm o}$ degrees, given by the solutions
$\vec{v}_1=r(1,\tanh{y})$, $\vec{v}_2=r(-\tanh{x},1),$
$\vec{v}_3=r(-1,-\tanh{y})$ and $\vec{v}_4=r(\tanh{x},-1)$.

Let us now show that there exist stable brane junctions. In order
for the system and its solutions that we have considered above to be
meaningful, we should slightly deform the space of parameter of the
theory \cite{blcag}. This is equivalent to consider the coupling
$\varepsilon$-term present in the Ginzburg-Landau equations above.
This in turn changes the Bogomol'nyi energy of the BPS solutions as
\cite{blcag} \ben\label{EbE}
E^{a}=E^{a}_B+\frac{1}{2}\varepsilon\int_{-\infty}^\infty F(u,v)dx,
\een where $E^{a}_B=|\Delta W_{a}|$ is the Bogomol'nyi energy and
$\varepsilon$ is the deformation parameter; the index $a=1,2,3,4$
indicates a particular edge, while $a=5,6$ indicates a particular
diagonal. The deformation $\varepsilon$-term which is compatible with
the Ginzburg-Landau equations anticipated in the introduction
above, can be inferred to be \ben\label{F}
F(u,v)=\frac{1}{2}(u^4+v^4)-3u^2v^2+\frac{9}{2}\,. \een Now we can
show that just one of the solutions (\ref{edge2}) or (\ref{diag2})
is stable while the other is unstable. In the limit $\varepsilon=0$
we find marginal stability of these solutions that also coincides
with the limit where the system decouples. We take the energy of the
solutions to guide ourselves in our study of stability. The total
energy (\ref{EbE}) of the solutions (\ref{edge2}) are the same in
any edge sector they involve but half of the energy of the diagonal
solutions (\ref{diag2}). In order for the junction to be stable, the
sum of energies of the two adjacent edges should be smaller than the
energy of any diagonal, i.e., the junction condition:
$\,E^a+E^{a+1}<E^b$ \cite{cht}. We compute the energies to get \ben
\label{EE}
E^{a}+E^{a+1}=4+\frac{21}{2}\varepsilon, \:\: a=1,2,3,4 \qquad {\rm edge}\\
E^{b}=4+6\varepsilon, \qquad b=5,6 \qquad {\rm 1-diag}
\een
Note that the four edge configuration is stable for $\varepsilon<0$, because in this regime it satisfies the junction condition. The
1-diag configuration, however, is stable for $\varepsilon>0$. Thus the configurations (\ref{edge2}) and (\ref{diag2}) cannot be
simultaneously stable. In any scenario involving $\varepsilon<0,$ domain walls (from the 1-diag sectors) separating adjacent vacua
tend to bifurcate into four half domain walls (from the edge sectors) joined together forming a 0-brane junction with one lower
dimension. This may represent a mechanism of relaxing dimensions \cite{kr2005} where a $p+1$-dimensional Universe initially formed
relaxes to another $(p-1)+1$-dimensional Universe, with one bifurcation.

{\it III. Three-field model.} ---The extension for three fields uses the superpotential
\ben
\label{W3}W=r\,(u+v+w)-\frac{1}{r}\left(\frac{u^3}{3}
+\frac{v^3}{3}+\frac{w^3}{3}\right).
\een
As in the former example, one-dimensional static soliton solutions can be found from the first-order equations
\ben\label{BPS3}
\frac{du}{dx}=r-\frac{u^2}{r},\:\:\:\:
\frac{dv}{dx}=r-\frac{v^2}{r},\:\:\:\:
\frac{dw}{dx}=r-\frac{w^2}{r}.
\een
The vacuum solutions are now given by $\bar{u}=\bar{v}=\bar{w}=\pm\,r$. These vacua are vertices of a cube in the space $(u,v,w)$. They comprises $2^3=8$ vacua and form a set of $28$ ($8!/2!\,6!$) independent BPS sectors, each one connecting a pair of vertices. These sectors compose $12$ $edges$, $12$ {\it surface diagonals} and $4$ {\it internal diagonals} of the cube, which we name edge, 1-diag, and 2-diag states. These sectors give us the following soliton (kink) solutions
\ben\label{edge3}
u_a^2=v_a^2=r^2;\:\:
w_a^2=r^2\,\tanh^2{(x)},\!\qquad\qquad\:{\rm edge}
\\
\label{diag3} u_a^2=r^2;\:\: v_b^2=w_b^2;\:\:
w_b^2=r^2\,\tanh^2{(x)},\: {\rm 1-diag}
\\
\label{intdiag3} u_c^2=v_c^2=w_c^2;\:\:
w_c^2=r^2\,\tanh^2{(x)},\:\:\qquad {\rm 2-diag} \een where
$a=1,...,12$, $b=13,...,24$ and $c=25,...,28$ label all the 28
sectors. Again, our point here is to show that in the case $\varepsilon\neq0$,
the system may {\it favor} the configuration of {\it
twelve edges} solutions (\ref{edge3}) instead of internal diagonal
solutions (\ref{intdiag3}). This is the analog of the bifurcation
phenomenon we have stressed above in the planar system. As we show
below, there also exist values of $\varepsilon$ which favor a
bifurcation where the {\it twelve edges configuration} is stable.
Such configuration can form stable junctions which is, in three
spatial dimensions, again a 0-brane junction. There would also exist
configurations of just 1-brane junctions coming from the 1-diag
sectors (\ref{diag3}), but they are not energetically favored. As we
can see below, the system prefer to bifurcate twice until the edge
configuration (\ref{edge3}) is achieved. Again, in the {\it
thin-wall limit} \cite{cht}, the junction can be represented by
$twelve$ half-walls rotated of $90^{\rm o}$ degrees, given by
solutions $\vec{v}_a=(u_a,v_a,w_a)$, with $a=1,...,12$. The first
three solutions are $\vec{v}_1=r(1,\tanh{y},1)$,
$\vec{v}_2=r(-\tanh{x},1,1)$, and $\vec{v}_3=r(-1,1,-\tanh{z})$.
Similar early arguments can be used to show that the twelve edge
configuration is really stable, provided we consider the function in
(\ref{EbE}) as $F(u,v,w)=F(u,v)+F(u,w)+F(v,w)$. The stability of the
junction is ensured if the sum of energies of the $three$ adjacent
edges is smaller than the energy of any 2-diag state, i.e.,
$E^a+E^{a+1}+E^{a+2}<E^c$. We compute the energies to get
\ben
\label{EEE} E^{a}+E^{a+1}+E^{a+2}=6+\frac{63}{2}\varepsilon,\:\: \:a=1,...,10 \:\:{\rm edge}\:\\
E^{b}+E^{b+1}=6+27\varepsilon, \:\:\: b=12,...,23 \:\:\:\:\:\:\:\:{\rm 1-diag}\:\\
E^{c}=6+18\varepsilon,\qquad \qquad c=25,...,28.\:\:\:\:\:\:\:\:\:{\rm 2-diag}\:
\een
Note that the twelve $edge$ configuration is $stable$ for $\varepsilon<0$, because in this regime it satisfies
the junction condition. Note that the 2-diag configurations have the largest energy, and then can be stable only
if $\varepsilon>0$. The 1-diag states that have intermediate energy decay in any case. Thus the configurations (\ref{edge3}),
(\ref{diag3}) and (\ref{intdiag3}) cannot be simultaneously stable. For our purposes, however, we assume $\varepsilon<0$ which leads us to
a scenario involving domain walls (from the 1-diag sectors) separating two adjacent vacua that tend to
bifurcate into twelve half domain walls (from the edge sectors) joined together forming a $three$ lower
dimensional 0-brane junction. For $n$ determining the number of scalar fields and the total number of bifurcations, these examples can naturally be extended to $n$ scalar fields, in $d$-spatial dimensions, to study stable junctions of $p$-branes to form $d-n$ dimensional brane junctions.

{\it IV. Einstein solid model.} ---In general in our model, the number of fields involved are $\geq$ the number of spatial
dimensions of space-time in which they live. As such, the smallest spaces are $0$-brane junctions which are zero dimensional. Since
gravity, matter and gauge fields tend to localize on the smallest $p$-brane junction, this seems to be a problem because things would
tend to collapse to isolated points. However, since the space-time is considered as an elastic system made out of $9n$ independent
one-dimensional oscillators, it is very natural to consider the model of an Einstein solid in nine spatial dimensions. Thus, at some
temperature below the {\it Einstein temperature} $T_E=\omega_0$ (we take $\hbar=c=k_B=1,$ and $\omega_0$ as the natural frequency), the
quantum effects effectively excite less degrees of freedom ({\it dofs}). Indeed, the number of {\it dofs} activated {\it
independently} as rotational and vibrational {\it dofs} of a diatomic molecule in a gas,  can be obtained from thermal effects,
by the {\it specific heat} of the system. In our Einstein solid, the independent {\it dofs} are the number of 8-branes and the mean
energy per 9Vol$_8$ is computed as $U(T)=\sum_{n=0}^N \epsilon_n\exp{(-\epsilon_n/T)}=T_8/[\exp{(T_8/T)}-1]$, with $\epsilon_n=nT_8$
and $N\!\to\!\infty$ is the total number of 8-branes. Here the ``quantum'' itself is the brane tension $T_8\!\equiv\!T_E$ of each
independently activated 8-brane. Let us now adopt the specific heat per Vol$_8$, $n(T)=9\,\partial U/\partial T$, as the effective
number of {\it dofs} varying with the temperature. This behaves as $n(T\!\gg\!T_E)\to 9$ while $n(T\!\ll\!T_E)\to0$; i.e., almost all
{\it dofs} are active (frozen) at very large (small) temperatures, but at intermediate temperatures only the thermally favorable {\it
dofs} activate. For $T\simeq(1/2)\, T_E$ the mean energy per 9Vol$_8$ approaches the zero point energy
$({1}/{2})\,\omega_0=({1}/{2})\,T_E$. Thus at such temperature, our
nine dimensional solid, oscillates at long wavelengths with an
effective number of {\it dofs} given by \ben n(T_E)\simeq6.48. \een
This signalizes that effectively only six {\it dofs} are indeed
active as the system is around the zero point energy. Thus, at such
regime, we can write an effective field theory for our nine
dimensional solid with only $six$ scalar fields, rather than $9n$
fields. We note that this mechanism reduces the spatial dimensions
to brane junctions with effectively $d-[n(T)]=3$ dimensions ([\;]
stands for integer part). This, of course, favors $3$-brane
junctions, made out of $[n(T)]$ codimension one branes intersecting
orthogonally. As we have already stated, a coupling
$\varepsilon$-term is required to get stable junctions. However,
such coupling can be infinitely small, so that we can use the
Einstein model of solid instead of a Debye model. A statistical
route for a system of $n$ copies of decoupled scalar fields has also
been employed in the study of a landscape of vacua in
Ref.~\cite{ADK}.

{\it V. Gravity, gauge and matter fields}. ---To show why a junction
is a natural space-time selection mechanism, let us first consider
the localization of gravity on $p$-branes and $p$-brane junctions
\cite{rs2,nima2000}; see also \cite{sg}. The brane junction solution
is achieved by considering $AdS_{10}$ slices glued together along
brane $\delta$-function sources. The spin-2 graviton field equations
can be found from the Lagrangian ${\cal L}_2\!=\!\sqrt{g}\,[{\cal
L}+(1/2\kappa^2)R+\Lambda]$, where $\cal L$ is given by
(\ref{action}). It is, however, important to show that the bulk
cosmological constant $\Lambda$ in our setup is indeed negative.
Using the fact that the scalar potential in (\ref{action}) is
perturbed by the $\varepsilon$-term of (\ref{EbE}) we find
$V=\frac12\sum_{i=1}^n(\partial_{u_i}W)^2+\frac12\varepsilon
F(u_1,u_2,...,u_n),$ where the function $F(u_1,u_2,...,u_n)$
represents $n$ fields combined in the form
$F(u_1,u_2,...,u_n)\!=\!F(u_1,u_2)+F(u_1,u_3)+...+F(u_{n-1},u_n)$
--- see Sec.~$III$ above. Now applying the explicit form of the
superpotential and $F(u_i,u_j)$ given above, we find that the
perturbed vacua are given by 
$\bar{u}_1\!=\!\bar{u}_2=\!...\!=\!\bar{u}_n\!=\!\pm
\!(3/[2-3(n-1)\varepsilon])^{1/2}$. Thus the bulk cosmological
constant defined as $\Lambda\!\equiv\!V(\bar{u}_1,\bar{u}_2,...,\bar{u}_n)$ reads
$\Lambda=-(27/8)n(n-1)^2\varepsilon^2/[2-3(n-1)\varepsilon],$ for $n>1,$
which is always $negative$ because as we have earlier stressed
$\varepsilon$ should be negative to stabilize the junctions. This result is crucial for the present
investigation, because it circumvents the former result of Ref.~{\cite{nag}}, which stresses
that brane junctions in a bulk flat space (where $\Lambda$ should vanish) are not generically able to
localize massless fields on them.

In the thin wall limit the ${\cal L}|_{kink}\to -\sigma
\sum_{I=1}^{6}{\delta(Z_I)}$, being $\sigma$ the 8-brane tension.
The fluctuations of the metric around the brane geometry
$ds_{10}^2=\Omega^{-2}(\eta_{\mu\nu}dX^\mu
dX^\nu+\sum_{I=1}^{6}{(dZ_I)^2)}$ are governed by a
Schr\"odinger-like equation \cite{rs2}. For a collection of six
$thin$ $8$-branes intersecting orthogonally we find \cite{nima2000}
\ben\label{schGraviton}
\left[-\frac{1}{2}\nabla^2+60k^2\Omega^2-4k\Omega\,\sum_{I=1}^{6}{\delta(Z_I)}
\right]\Psi=m^2\Psi. \een The zero mode is confined to the junction
whose wave function is
$\Psi_0(Z)\!=\!C\Omega^4\!=\!C/{(\sum_{I=1}^{6}{k|Z_I|+1})^{4}}.$
Note that the zero mode probability density $|\Psi_0(Z)|^2$ is
maximized (or ``peaked'') as $|Z_1|+|Z_2|+...+|Z_6|=0$, i.e., at
$Z_1,...,Z_6\!=\!0$. This defines a $3$-brane junction in ten
space-time dimensions. Note we also have other junctions such as
``edges'', e.g., a $7$-brane junction, but the zero mode is not
peaked there. Thus, as we have anticipated, gravity ``prefers'' to
live on the $3$-brane junction, i.e., the lower dimensional
junction. The massive graviton states can be also localized on
$p$-branes or $p$-brane junctions. However, the corrections to the
Newton potential related to such massive states are highly
suppressed at large distances. Thus, at large distances, $p+1$
dimensional gravity on $p$-branes or $p$-brane junctions is
correctly described just by the graviton zero mode
\cite{rs2,nima2000}.

Similar ideas have been recently used in \cite{R} to localize gravity on triple intersection of 7-branes in 10 dimensions.

The presence of the six active background fields describing the kink solutions above favors localization of matter and gauge fields
\cite{add}. To construct a field theory with localization of spin-1 fields, one can use the coupling $\sum_I(\partial\Phi_I)^2F^2$,
which is a term in the expansion of the Born-Infeld action for a 3-brane, and Yukawa couplings $\sum_{IJ}\partial_I\partial_J
W(\Phi)\bar{\Psi}_I\Psi_J$ for spin-1/2 fields. In the thin wall limit, $(\partial\Phi_I)^2|_{kink}\to\sigma\delta(Z_I)$ implies the
gauge fields get confined, while the fermion mass matrix $\partial_I\partial_J W(\Phi)|_{kink}\!\to\!\delta_{IJ}\,{\rm
sgn}{(Z_I)}$ approaching zero at $Z_1,...,Z_6\!=\!0$ favors the localization of fermion zero modes. They suggest the localization of
$six$ copies of fermions on a $3$-brane junction. They could be the less massive states forming the {\it three families} of quarks and
leptons observed in our world.

The authors would like to thank CAPES, CNPq, PADCT/CNPq, and PRONEX/CNPq/FAPESQ for partial financial support.

\end{document}